\documentclass[10pt,draft]{dis03}
\usepackage{epsf,amsmath}

\newcommand{\BQ}{\begin{equation}}
\newcommand{\EQ}{\end{equation}}
\newcommand{\BQA}{\begin{eqnarray}}
\newcommand{\EQA}{\end{eqnarray}}
\newcommand{\be}{\begin{eqnarray}}
\newcommand{\ee}{\end{eqnarray}}

\newcommand{\x}{x_\perp}
\newcommand{\y}{y_\perp}

\newcommand{\rr}{r_\perp}

\newcommand{\kk}{k_\perp}

\def\simge{\mathrel{%
   \rlap{\raise 0.511ex \hbox{$>$}}{\lower 0.511ex \hbox{$\sim$}}}}
\def\simle{\mathrel{
   \rlap{\raise 0.511ex \hbox{$<$}}{\lower 0.511ex \hbox{$\sim$}}}}

\begin{document}

\title{}\author{}
\maketitle
\vspace*{-25mm}

\begin{flushright}
{\normalsize SACLAY--T03/135}\\
\end{flushright}

\begin{center}
{\bf Deep Inelastic Scattering in the Color Glass Formalism
\footnote{Contribution to the 11th International Workshop on Deep-Inelastic Scattering (DIS03), St. Petersburg, April 2003.}}\\

\vspace*{2mm}

Kazunori Itakura\\
\vspace*{2mm}

RIKEN-BNL Research Center, BNL, Upton, NY 11973, USA,\\
SPhT CEA/Saclay, 91191, Gif-sur-Yvette Cedex, France\footnote{Present address}
\end{center}

\vspace*{-1cm}
\begin{abstract}
\noindent I discuss some phenomenological consequences of the 
Color Glass formalism for deep inelastic scattering.
\end{abstract}

\vspace*{-5mm}
\section{Dipole scattering at small $x$}
\vspace*{-3mm}
The Color Glass Condensate (CGC) is the effective theory for 
small $x$ gluons derived from QCD \cite{Edmond_Raju,Iancu_dis}. 
Since the theoretical framework of CGC has been established, 
a considerable progress has been made in understanding its properties. 
In this talk, I will discuss applications of the CGC formalism to 
the calculation of deep inelastic scattering (DIS). 
The relevant quantities to be discussed are the $\gamma^* p$ total cross 
section $\sigma_{\rm total}^{\gamma^*p}$ and the $F_2$ structure 
function. At small $x$, one can compute them as follows:
\vspace*{-2mm}
\BQA
\label{sigmaDIS}
\sigma_{\rm total}^{\gamma^*p}(x,Q^2)=
\sum_{\rm T,L}
\int_0^1 dz \int d^2r_\perp\,|\Psi_{\rm T,L}(z,r_\perp;Q^2)|^2\,
\sigma_{\rm dipole}(x,r_\perp),
\EQA
\vspace*{-8mm}
\BQA
F_2(x,Q^2)=({Q^2}/{4\pi^2 \alpha^{}_{\rm EM}})\  
\sigma^{\gamma^* p}_{\rm total}(x,Q^2),
\EQA
where $\Psi_{\rm T/L}(z,r_\perp;Q^2)$ are the LC wavefunctions 
of transverse/longitudinal virtual photons splitting into a 
$q\bar q$ dipole ($r_\perp$: transverse size,  
 $z$: a fraction of the photon's longitudinal momentum 
carried by the quark), and $\sigma_{\rm dipole}(x,r_\perp)$ 
is the {\it dipole-proton cross section} given by 
( $r_\perp\!\! =\! x_{\perp}-y_{\perp},$ 
$b_\perp\! =\!(x_{\perp}+y_{\perp})/2$ ) 
\vspace*{-2mm}
\be\label{sigmadipole}
\sigma_{\rm dipole}(x,r_\perp)\! = 2\! \int\! d^2b_\perp
(1-S_x(x_{\perp},y_{\perp})),\   
S_x(x_{\perp},y_{\perp})\equiv \frac{1}{N_c}
\left\langle {\rm tr}\big(V^\dagger_{x_{\perp}} V_{y_{\perp}}\big)
\right\rangle_{x}.\!\!\!
\ee
The average of the product of Wilson lines $V^\dag_{\x}V_{\y}$ 
is computable in CGC (see \cite{Edmond_Raju} for details).  
Therefore, once we obtain $\sigma_{\rm dipole}$ based on 
CGC, we can immediately calculate $\sigma_{\rm total}^{\gamma^*p}$ 
and $F_2$ from the formulae (1) and (2).
Below we explain three important
features of $\sigma_{\rm dipole}$ as computed within CGC.\\
\vspace*{-3mm}

\noindent{\bf (I) Geometric scaling persists beyond $Q_s^2$ up to 
      $Q^2\sim Q_s^4/\Lambda_{\rm QCD}^2$ \cite{IIM}.}  \\
Geometric scaling is a new scaling phenomenon at small $x$ \cite{GS}
meaning that $\sigma_{\rm total}^{\gamma^* p}$ depends upon 
$Q^2$ and $x$  only via their specific combination 
$\xi\equiv Q^2 R_0^2(x)$, with $R_0^2(x) \propto x^{\lambda}$. 
This can be naturally explained by CGC at low $Q^2$, 
below the {\it saturation scale} $Q_s^2$ ($\sim$  a few GeV$^2$). 
For example, the energy dependence of the saturation scale 
is found to be $Q_s^2(x)\propto x^{-\beta}$, which is 
consistent with that of $1/R_0^2(x)$.
The geometric scaling up to much higher values of $Q^2$ 
($\sim$ 100 GeV$^2$) can also be understood within the framework of 
CGC through the solution to the BFKL equation 
subjected to a saturation boundary condition at $Q^2\sim Q_s^2(x)$. 
Indeed, the solution exhibits approximate scaling 
\be
1-S_x(\rr)\simeq \left(\rr^2 Q_s^2(x) \right)^\gamma,
\ee
within a window  $1 \simle \ln(Q^2/Q_s^2) \ll 
\ln(Q_s^2/\Lambda^2_{\rm QCD})$, which is roughly 
consistent with phenomenology. 
The power $\gamma$ (more precisely, the difference $1-\gamma$) is 
called {\it anomalous dimension} since it is different from the 
naive DGLAP value $\gamma=1$ (i.e., $1-S_x(\rr)\propto \rr^2 $). 
Namely, the LO BFKL equation 
yields $\gamma\simeq 0.64$ \cite{IIM}, which does not change 
even with the higher order effects included \cite{Dionysis}.\\

\vspace*{-3mm}

\noindent{\bf (II) $\sigma_{\rm dipole}$ can be analytically 
computed within CGC \cite{Gaussian}.}\\
The evolution in CGC is described by a 
renormalization group equation for the weight function
which governs the correlation of color sources representing 
large $x$ partons. One can solve it approximately 
in two different kinematical regimes \cite{IM}, 
namely, in the weak field regime (low density, no saturation)
and in the strong field regime (deeply at saturation). 
The first regime is relevant for the scattering of a small dipole 
($\rr\ll 1/Q_s(x)$), while the second one applies for a relatively 
large dipole ($\rr\gg 1/Q_s(x)$).
In both regimes, the weight function can be represented 
as a Gaussian.
In between them, one can use a simple 
interpolation to get a Gaussian effective theory 
which may be used globally \cite{Gaussian}.
In particular, one can use this theory to deduce an 
explicit analytical form of the dipole $S$-matrix valid 
in the scaling region,  $Q^2=1/\rr^2 
\simle Q_s^4(x)/\Lambda_{\rm QCD}^2$:
\be
S_x(\rr)=\exp\left\{
-\kappa \int \frac{d^2\kk}{(2\pi)^2}
\frac{1-{\rm e}^{i\kk\cdot \rr}}{\kk^2}
\ln \left[\ 
1+\left(
\frac{Q_s^2(x)}{\kk^2}
\right)^{\gamma\ } 
\right]
\right\},
\ee
where 
$Q_s^2(x)\propto x^{-c\bar\alpha_s}$
and $\kappa$ is a known numerical constant \cite{Gaussian}.
This formula reproduces (4) in the extended scaling regime 
and has the correct behavior in the saturation regime 
$S_x(\rr)\propto \exp \{-\frac{1}{2c}
\ln^2 \rr^2 Q_s^2(x)\}$ \cite{IM}.\\

\vspace*{-3mm}

\noindent{\bf (III) $\sigma_{\rm dipole}(x,\rr)$ saturates 
 Froissart bound at high energy \cite{FB}.}\\
Consider the high energy scattering of a small $q\bar q$ dipole 
off a proton. With increasing energy, a {\it black disk} (BD) 
in which the dipole is strongly absorbed ($S_x(\rr,b_\perp)\simeq 0$) 
appears near the center of the proton, where the gluon density is larger,
and then expands over the target.
At high energy, $\sigma_{\rm dipole}\simeq 2\pi R^2(x,1/\rr^2)$, 
so we need to compute 
the growth of the BD radius $R(x,1/\rr^2)$ with decreasing $x$.
This was done in Ref.~\cite{FB} under the following assumptions:
(i) At impact parameters $b_\perp$ outside of (but not too far away from) 
    the BD, the gluon density  is relatively low (but still perturbative), 
    and grows rapidly with $1/x$ due to BFKL evolution.
    The dipole scattering is then dominated by relatively nearby color
    sources, within a disk of size $1/Q_s(b_\perp)$ around $b_\perp$. 
    (Sources which lie further away contribute less because of 
     color screening at saturation \cite{Gaussian,FB}.)
(ii) With increasing energy at fixed $b_\perp$, the local $S$-matrix
    decreases, and eventually becomes negligible 
    ($S_x(\rr,b_\perp)\simeq 0$),
    because of {\it saturation}. This corresponds to the unitarity limit 
    $T\simeq 1$ for the scattering amplitude $T_x(\rr,b_\perp)\equiv 1-S$. 
    When this happens, the point $b_\perp$ enters the BD.
(iii) 
   For sufficiently large energies, the BD enters the tail of the hadron 
    wavefunction, where the gluon density decreases exponentially with 
   $b_\perp$. Then the expansion of the BD is controlled by the competition 
   between the BFKL {\it growth} of the gluon density with $1/x$ and its 
   exponential {\it decrease} with $b_\perp$: $T_x(\rr,b_\perp)\propto 
   {\rm e}^{\omega \bar\alpha_s \ln 1/x}\, {\rm e}^{-\mu b}$, 
   with $\mu$ being the 
   lightest mass gap in the correct channel for dipole-hadron scattering.
Then the BD radius is found to be proportional to $\ln (1/x)$ 
which yields the result saturating  Froissart bound: 
\vspace*{-2mm}
\be
\sigma_{\rm dipole}(x,\rr)\simeq \frac{\pi}{2}
\left(\frac{\omega\bar\alpha_s}{m_\pi}\right)^2
\ln^2 \frac{1}{x}\ .
\ee
Here we took $\mu=2m_\pi$ with $m_\pi$ the pion mass.

Whereas points (i) and (ii) above refer to the perturbative (yet non-linear)
physics which is proper to the CGC formalism (and, in particular, also to
the Balitsky-Kovchegov (BK) equation \cite{BK}), on the other hand, 
point (iii) should be seen as a consequence of confinement, which 
is not encoded in the present formalism. A purely perturbative evolution 
based on the BK equation leads to the violation of Froissart bound 
\cite{KW,Stasto} because of long-range interactions which, in real 
world, are removed by confinement \cite{FB}. In fact, in Ref.~\cite{FB},
the result (6) has been obtained by following the perturbative evolution with 
non-perturbative initial conditions over a {\it limited} energy interval,
and by arguing that, in the presence of confinement,
the same result would hold up to arbitrary high energies.

\section{Phenomenological consequences \cite{Preparation}}
\vspace*{-3mm}
Now we use all the features (I) (II) and (III) to compute 
the physical quantities (1) and (2). First of all, let us 
give a rough estimate of $\sigma_{\rm total}^{\gamma^*p}$ 
with some approximations. We assume transverse homogeneity 
of the proton, ignore quark mass and the longitudinal cross section 
(which is much smaller than the transverse one), and use 
scaling form (4) for $S_x(\rr)$ in the extended scaling window, while 
simply $S_x(\rr)\simeq 0$ 
in the saturation regime. Then we obtain the following leading 
behaviors (numerical coefficients are suppressed)
\vspace*{-1mm}
\BQA
\sigma_{\rm total}^{\gamma^*p}(x,Q^2)\ \sim\ 
\left\{\begin{array}{lc}
\ln (Q_s^2(x)/Q^2) & Q^2<Q_s^2 \\ 
(Q_s^2(x)/Q^2)^\gamma       &
Q^2_s\simle Q^2 \simle Q_s^4/\Lambda_{\rm QCD}^2
\end{array}
\right. .
\EQA
Note first that the scaling in $S_x(\rr)$ (w.r.t. $\rr^2Q_s^2$) 
has been converted into the experimentally observed scaling 
w.r.t. $Q^2/Q_s^2(x)$.
Next, interestingly, this result is consistent with the experimental 
data while this is admittedly a very rough estimate. In particular, 
the slope of the data plotted against $Q^2/Q_s^2(x)$
(in log-log scale) is in rather good agreement with the BFKL value 
$\gamma\simeq 0.64$ in the extended scaling window.

In (II), we have shown a formula (5) for the $S$-matrix.
This immediately gives the dipole cross section 
$\sigma_{\rm dipole}^{\rm CGC}(x,\rr)=2\pi R^2 [1-S_x(\rr)]$ with $R$ 
being the hadron radius. 
In fact there are other models for the dipole cross section which
incorporate the effects of saturation (for references, see 
\cite{Edmond_Raju}).  
The first one is by Golec-Biernat and W\"usthoff 
$\sigma^{\rm GBW}_{\rm dipole}(x, \rr)=\sigma_0 [1-\exp\{-\rr^2 Q_s^2(x)/4\}]$
with $Q_s^2=$GeV${}^2(x/x_0)^\lambda$, and has only three parameters 
($\sigma_0,  x_0, \lambda$). 
This gives the naive value for the exponent $\gamma=1$. 
The second one is by Bartels, Golec-Biernat and Kowalski
$\sigma_{\rm dipole}^{\rm BGK}(x, \rr)=\sigma_0 
[1-\exp\{- (\pi^2 \alpha_s/\sigma_0)\rr^2 xG(x,\mu^2)\}]$ 
having five parameters ($\sigma_0,$ two for $\mu^2$ and two for
the initial condition of $xG$). This includes the DGLAP evolution
and thus improves significantly over the GBW model at high $Q^2$.
These simple parametrizations were quite successful in fitting HERA data. 
As compared with such previous parametrizations, our formula reproduces 
the correct QCD behaviors in both the saturation and extended scaling regimes.
Nevertheless, our purpose with eq.~(5) is not to provide another fit
--- this equation is rather crude, and is expected to apply
only in a limited kinematical regime which is only marginally explored 
at HERA ---, but rather to use this estimate for $\sigma_{\rm dipole}$ 
to draw some qualitative conclusions, rooted in QCD, about the 
expected behavior of the DIS structure functions at sufficiently small $x$.
One of these conclusions is summarized in eq.~(7). Other results of this 
type will be presented somewhere else \cite{Preparation}.
Here, we shall briefly comment on one of these conclusions, 
related to the high energy limit of $F_2$. 
Note that the result (6) comes from the leading order behavior 
of the BD at small $x$ for fixed dipole size $\rr$. 
Since $F_2$ is given by the 
integration over various size of the dipoles, we need to consider also
the dipole size dependence of the BD radius $R(x,1/\rr^2)$, which is given
in \cite{FB}. Then, using the dipole cross section 
$2\pi R^2(x,1/\rr^2)$, one can compute $F_2$ in the high energy limit.
The results is $F_2(x,Q^2)\propto \ln^3 (1/x)$, and thus 
$F_2$ and $\sigma_{\rm total}^{\gamma^*p}$ appear to violate 
Froissart bound. This is not a problem because 
the virtual photon is not a (properly normalized) hadronic state, 
but rather a superposition of such states (one for each dipole size).
The additional factor of $\ln (1/x)$ comes from integrating over all 
dipole states with the photon wavefunction \cite{Preparation}.


\end{document}